\documentclass[aps,prb,twocolumn,showpacs]{revtex4}
\usepackage{amsmath}

\usepackage{graphicx}

\begin{document}

\title{Flux dynamics and vortex phase diagram in Ba(Fe$_{1-x}$Co$_x)_2$As$_2$
single crystals revealed by magnetization and its relaxation}

\author{Bing Shen, Peng Cheng, Zhaosheng Wang, Lei Fang, Cong Ren, Lei Shan and Hai-Hu Wen\email}

\affiliation{National Laboratory for Superconductivity, Institute of
Physics, Chinese Academy of Sciences, P.~O.~Box 603, Beijing 100190,
P.~R.~China}
\date{\today}

\begin{abstract}
Magnetization and its relaxation have been measured in
Ba(Fe$_{1-x}$Co$_x)_2$As$_2$ single crystals at various doping
levels ranging from very underdoped to very overdoped regimes.
Sizable magnetization relaxation rate has been observed in all
samples, indicating a moderate vortex motion and relatively small
characteristic pinning energy. Detailed analysis leads to the
following conclusions: (1) A prominent second-peak (SP) effect  was
observed in the samples around the optimal doping level ($x \approx$
0.08), but it becomes invisible or very weak in the very underdoped
and overdoped samples; (2) The magnetization relaxation rate is
inversely related to the transient superconducting current density
revealing the non-monotonic field and temperature dependence through
the SP region; (3) A very sharp magnetization peak was observed near
zero field which corresponds to a much reduced relaxation rate; (4)
A weak temperature dependence of relaxation rate or a plateau was
found in the intermediate temperature region. Together with the
treatment of the Generalized-Inversion-Scheme, we suggest that the
vortex dynamics is describable by the collective pinning model.
Finally, vortex phase diagrams were drawn for all the samples
showing a systematic evolution of vortex dynamics.
\end{abstract}


\maketitle

\section{Introduction}
Since the discovery of superconductivity at $T_c$ = 26
K\cite{Kamihara} in LaFeAsO$_{1-x}$F$_ x$, the iron-based layered
superconductors have exposed an interesting research area to the
community of condensed matter physics. This breakthrough was
followed by the realization of superconductivity at temperatures at
as high as 56-57 K in Ca$_{1-x}$REFeAsF (fluorine derivative
¡°1111¡± system with RE = Nd, Sm, etc.).\cite{ChengPEPL2009} A lot
of experimental and theoretical works on the physical properties
have been carried out. For the FeAs-1111 phase, it is very difficult
to grow crystals with large sizes, therefore most of the
measurements so far were made on polycrystalline samples. This is
much improved in the (Ba,Sr)$_{1-x}$K$_x$Fe$_2$As$_2$ and
(Ba,Sr)(Fe$_{1-x}$Co$_x)_2$As$_2$ (denoted as FeAs-122) system since
sizable single crystals can be produced.\cite{Johrendt122} Specific
heat, lower critical field, microwave measurements suggest that
these superconductors exhibit multiband feature
\cite{HunteF,RenC,Machida1}. Measurements under high magnetic fields
reveal that the iron-based superconductors have very high upper
critical fields,\cite{ HunteF,Senatore,JiaY} which indicate
encouraging potential applications. Preliminary experimental results
also indicate that the vortex dynamics in iron pnictide may be
understood with the model of thermally activated flux motion within
the scheme of collective vortex
pinning.\cite{YangHPRB,YangHAPL,ProzorovPRB,ProzorovPhysicaC} A
second-peak (SP) effect has been observed in
Ba$_{1-x}$K$_x$Fe$_2$As$_2$\cite{YangHAPL} and
Ba(Fe$_{1-x}$Co$_x)_2$As$_2$ single
crystals.\cite{ProzorovPRB,Prozorovarxiv, Y. Machida,HARDY V}
Moreover this system has a layered structure with the conducting
FeAs layers being responsible for the superconductivity, which
behaves in a similar way as the case of cuprates. For cuprate
superconductors, due to the high anisotropy, short coherence length,
and high operation temperature, the vortex motion and fluctuation
are quite strong.\cite{Blatter} This leads to a small characteristic
pinning energy, and the single vortex or vortex bundles are pinned
collectively by many small pinning centers. Therefore it is curious
to know whether the vortex properties and phase diagram of the
cuprate and FeAs-based superconductors are similar to each other or
not.\cite{Y.Yeshurun,Brandt,Beek} In this paper, we report an
intensive study on the vortex dynamics of
Ba(Fe$_{1-x}$Co$_x)_2$As$_2$ single crystalline samples ranging from
very underdoped to very overdoped regime.

\section{Experiment}

The single crystals with high quality  measured in this paper were
prepared by the self-flux method.\cite{LFang} Samples with six
different doping concentrations ($x$ = 0.06, 0.07, 0.08, 0.1, 0.12,
0.15) with typical dimensions of $1.0 \times 1.0 \times0.3$ mm$^3$
have been used for both magnetic and resistive measurements. The
measurements were carried out on a physical property measurement
system (PPMS, Quantum Design) with the magnetic field up to 9 T.
During the measurements, the magnetic field H is always parallel to
c-axis of single crystals. The temperature stabilization of the
measuring system is better than $0.01$ K. The magnetic properties
were measured by the sensitive vibrating sample magnetometer at the
vibrating frequency of 40 Hz with the resolution better than $1
\times10^{-6}$ emu. The magnetic field sweeping rate can be varied
from 0.5 Oe/s to 627 Oe/s, for most of the measurements of dynamical
magnetization relaxation measurements (defined below) we adopted the
sweeping rate of 50 Oe/s and 200 Oe/s, which were much enough for us
to resolve the difference of magnetization. The advantage of this
technique is that the speed of data acquisition is very fast with a
quite good resolution for magnetization.

 \begin{figure}
\includegraphics[width=8cm]{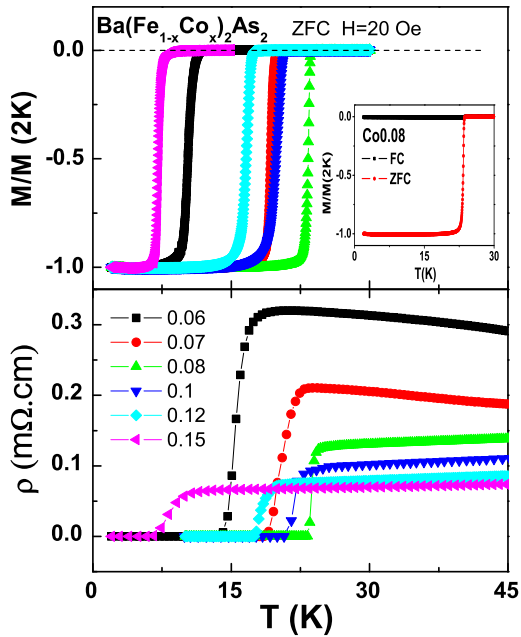}
\caption{ (Color online) Temperature dependence of the
superconducting diamagnetic moment (upper panel) and resistivity
(bottom panel) of six doped samples. The inset of upper panel shows
the temperature dependence of the diamagnetic moment measured in the
ZFC and FC processes at a field of 20 Oe for the sample $x$ = 0.08.
The sharp magnetization and resistivity transition curves assure the
high quality of the samples.}\label{fig1}
\end{figure}

In Fig.1 we show the temperature dependence of the diamagnetic
moment and resistive transitions of the six doped samples. The
superconducting transition temperature $T_c$ is found to shift
systematically with doping, revealing a dome-like doping dependence
(shown later in Fig.20). The sample with nominal composition $x$ =
0.08 was found to be optimally doped with the highest onset
transition temperature $T_c\approx 24.5 K$. In the underdoped region
($x < 0.08$), an upturn in the resistivity curve above $T_c$ can be
easily seen, which was supposed to be related to the structural and
antiferromagnetic (AF) transition. The large difference between zero
field cooling (ZFC) and field cooling (FC) magnetization (shown in
inset of Fig.1) indicates a strong magnetization hysteresis in the
sample. The perfect diamagnetism in the low temperature region and
sharp transitions observed from the ZFC curves indicate the high
quality of our samples.
\begin{figure*}

\includegraphics[width=22cm]{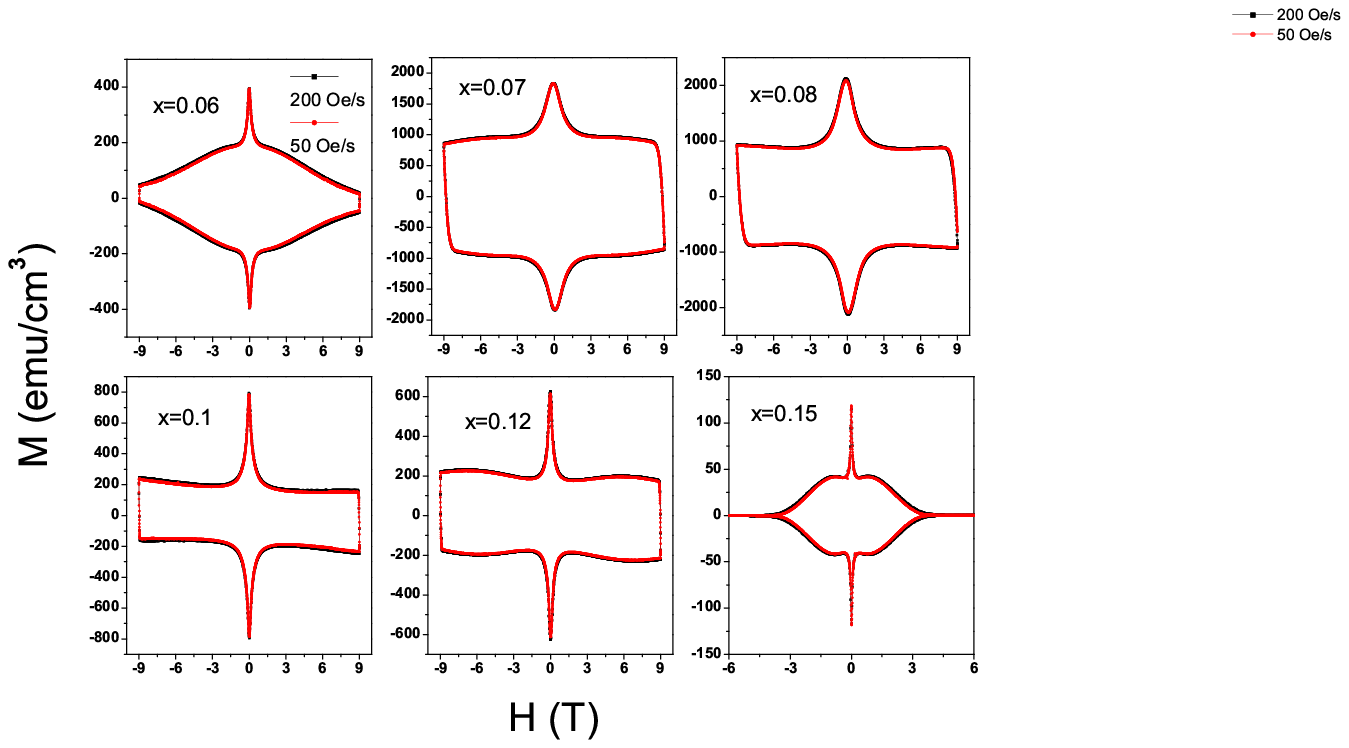}
\caption{(Color online)Magnetic hysteresis loops $M$ vs $H$ of the
six doped samples at 2 K. All the samples were measured with the
field sweeping rate of 50 Oe/s (shown by red circles) and 200 Oe/s
(shown by black squares). One can see that the second-peak or
fish-tail effect appears near the optimal doping, while it becomes
invisible in the very underdoped and very overdoped samples.
}\label{fig2}
\end{figure*}

To investigate the vortex dynamics, in this work we adopt the
so-called dynamical relaxation
method.\cite{Jirsa,WenPhysicaC1995,WenPRB1995} Dynamic
magnetization-relaxation measurements were carried out in the
following way:  The sample is cooled down to a chosen temperature in
ZFC mode and then we sweep the magnetic field and measure the
magnetic moment by following the routes: $0\rightarrow
H_{max}\rightarrow 0\rightarrow -H_{max}\rightarrow 0\rightarrow
H_{max}$ with different field sweeping rates d$B/$d$t$. The
corresponding magnetization-relaxation rate $Q$ is defined as

\begin{equation} Q \equiv \frac{\textrm{d} \ln j_s}{\textrm{d} \ln (dB/dt)}=\frac {\textrm{d} \ln
(\Delta M)}{\textrm{d} \ln(\textrm{d}B/\textrm{d}t)} ,
\end{equation}
where $j_s$ is the transient superconducting current density,
$\textrm{d}B/\textrm{d}t$ is the field sweep rate. Comparing to the
conventional magnetization-relaxation measurement (fixing the
magnetic field and measuring the time dependence of magnetization),
this dynamical relaxation method can overcome the following
drawbacks: (1) For the samples with a large demagnetization factor,
a slight overshoot of the field (even lower than 1 mT) can modify
the current distribution dramatically;(2) A long waiting time is
necessary before meaningful relaxation data points can be recorded;
(3) To get the valid relaxation data, we need to measure the
magnetization relaxation in a long time period.

In Fig. 2 we show the magnetization hysteresis loops (MHL) of six
doped samples measured at 2 K with the magnetic sweeping rates of 50
Oe/s and 200 Oe/s, respectively. The symmetric MHL curves indicate
that the bulk superconducting current instead of the surface
shielding current dominates in the samples during the measurements.
A surprising observation here is that the difference between $M$
measured at 200 and 50 Oe/s can be easily distinguished, which
indicates a relatively large vortex creep rate, or as called as
giant vortex creep in the cuprate superconductors. A prominent SP
effect  or called as the fish-tail effect was observed in the
samples around the optimal doping level ($x$ = 0.08), but it becomes
invisible in the very underdoped sample($x$ = 0.06) and hardly
visible in the highly overdoped samples ($x$ = 0.15). When the field
is approaching zero, the absolute value of magnetic moment increases
markedly. The central peaks are surprisingly sharp in all the six
doped samples which was hardly seen in the conventional
superconductors and high temperature superconductors (HTSC). Whereas
in FeAs based superconductors, this kind of sharp peaks are often
observed in both 1111 and 122 systems, which will be detailed in
next section.\cite{YangHPRB,YangHAPL,ProzorovPRB,HARDY V}

\section{Results and Data Analysis}

\subsection{The optimally doped sample}

\begin{figure}
\includegraphics[width=8cm]{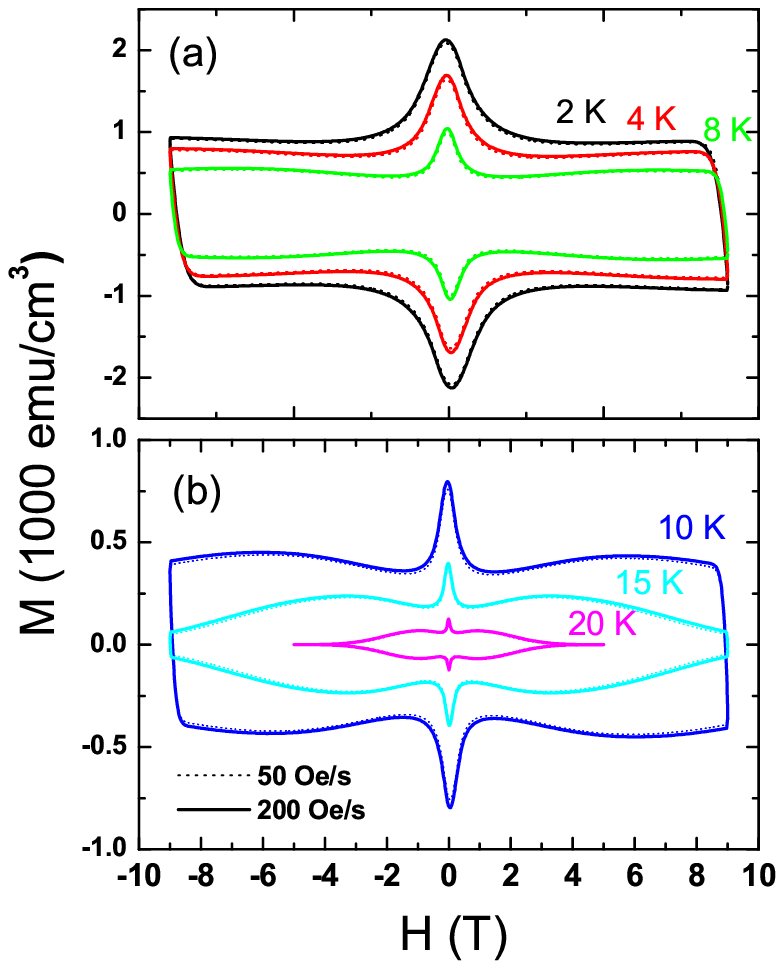}
\caption{(Color online) The MHLs of the optimally doped sample
measured at 2 K, 4 K, 8 K, 10 K, 15 K and 20 K. The solid line was
measured at the magnetic field sweeping rate of 200 Oe/s while the
dotted line at 50 Oe/s. A second-peak effect appears in all the
MHLs. }\label{fig3}
\end{figure}

\subsubsection{The sharp central peak and second-peak}
The MHLs measured at different temperatures from 2 K to 20 K are
presented in Fig.3 for the optimally doped sample ($x$ = 0.08). The
symmetric curves suggest that the bulk pinning instead of the
surface barrier dominates in the sample. The SP effect can be easily
observed in Fig.3(b) at $T >$ 10 K. With the decreasing of
temperature, the SP moves to higher magnetic field and finally goes
beyond the maximum field value 9 T as shown in Fig.3(a). The global
shape of MHL and the related features resemble that in the cuprate
superconductor YBa$_2$Cu$_3$O$_{7-\delta}$.\cite{SPYBCO}

\begin{figure}
\includegraphics[width=8cm]{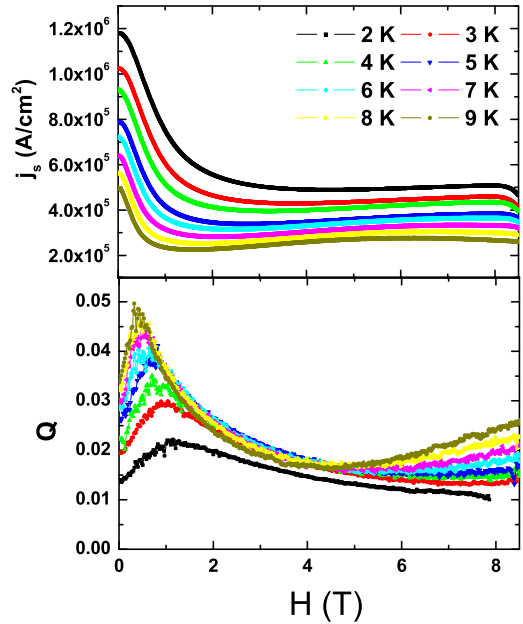}
\caption{(Color online)(Upper panel) Magnetic field dependence of
the MHL width $\bigtriangleup M$ and (Lower panel) dynamic
magnetization relaxation rate from 2 K to 9 K with the sample of $x$
= 0.08. The SP position corresponds to a minimum of the
magnetization relaxation rate Q. Near zero field, a sharp drop of
relaxation rate was observed in accompanying with the sharp central
peak of MHL.}\label{fig4}
\end{figure}

\begin{figure}
\includegraphics[width=8cm]{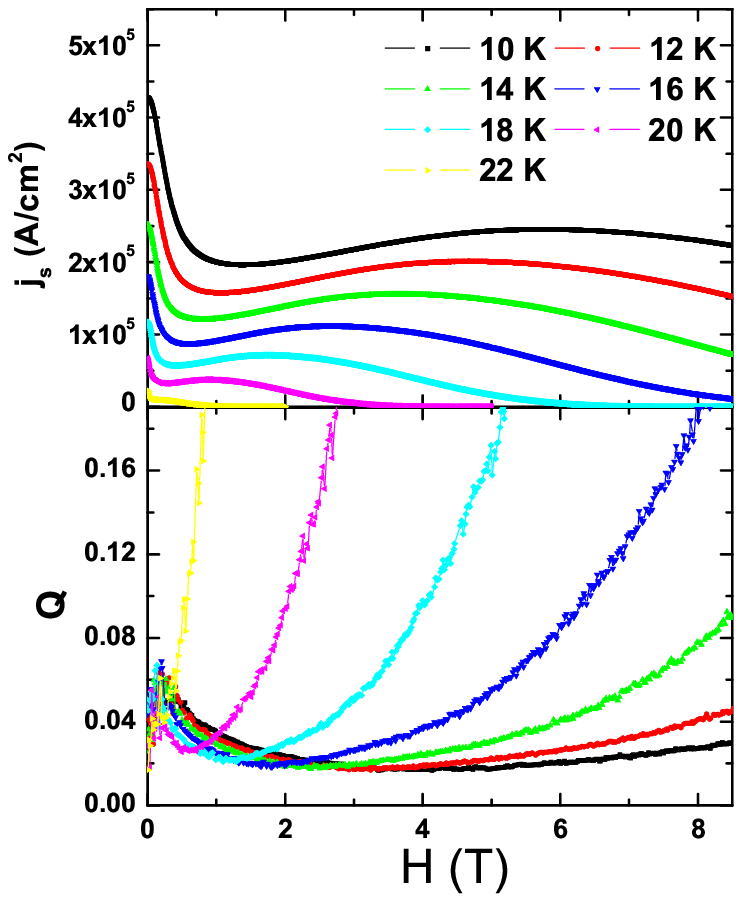}
\caption{(Color online) Magnetic field dependence of the transient
superconductivity current $j_s$ and (Lower panel) dynamic
magnetization relaxation rate from 10 K to 22 K with the increments
of 2 K for the sample of $x$ = 0.08. The SP position corresponds to
a minimum of the magnetization relaxation rate $Q$. Near zero field,
a sharp drop of relaxation rate was observed in accompanying with
the sharp central peak of MHL. A quick and monotonic rising of the
relaxation rate after the SP position was observed. }\label{fig5}
\end{figure}

When the external field is approaching zero, the absolute value of
magnetic moment increases markedly and reaches a maximum, showing a
very sharp magnetization peak near zero field. In high temperature
superconductors (HTSC) and other conventional type-II
superconductors, we can see a magnetization hump near zero field.
This hump near zero field was illustrated by mathematical simulation
as due to the following reason:\cite{H.G. Schnack} The low flow rate
or strong pinning in the interior part prevents the vortices from
leaving the sample rapidly, while when the external field is
approaching or just departing from zero, the vortices of opposite
signs enter the sample, which makes the number of vortices decrease
quickly at the edge. This produces a non-linear $B(x)$ profile with
a high slope near the edge. Clearly, in our six doped samples the
central magnetization peaks are surprisingly sharp. This sharp
magnetization peak near zero field in iron pnictide may be
understood in the similar way. When the external field is swept back
to zero, because of the small absolute value of $B(x)$, the slope of
$B(x)$ and the critical current density near the edge is much larger
than that in the interior part. In this case, a much enhanced
magnetization will appear. However, this simple picture will be
complicated by the step of $B(x)$ in the surface layer with
thickness of penetration depth. Meanwhile the surface barrier and/or
the geometrical barrier may also play roles here.

Based on the Bean critical state model,\cite{C. P. Bean} we calculate the
superconducting current density
 \begin{equation}
j_s=20\frac{\Delta M}{w(1-\frac{w}{3l})},
\end{equation}
where $\Delta M = M_{+}-M_{-}$, and $M_{+}$ ($M_{-}$) is the
magnetization associated with increasing (decreasing) field; $w, l$
is the width and length of the sample separately. In Fig.3 and
Fig.4, we present the field dependence of $j_{s}$ and relaxation
rate $Q$. As shown in Fig.3 and Fig.4, the magnetization relaxation
rate is inversely related to the $j_{s}$ revealing the non-monotonic
field and temperature dependence through the SP region. Therefore
the SP effect is induced by the transient relaxation effect which
depends on the time scale.\cite{LiSLPRB,Yeshurun} As mentioned
above, near zero field, a clear sharp magnetization peak can be
found in MHL. Accompanying this peak, there is a clear suppression
of the relaxation rate which exhibits a valley near zero field. This
may be related to the stronger critical current density in the edge
region where the $B(x)$ value is small. With increasing field, a
peak of magnetization relaxation can also be observed region
corresponding to the crossover of $j_s$ between the low-field high
slope and high-field low slope. This may suggest that the crossover
point (or the peak position of $Q$) corresponds to two different
regimes of vortex dynamics. The position of the peak shift to the
lower field with the increasing the temperature. A full
understanding to this effect would need a local measurement
facility, such as Hall probe array and magneto-optics. We leave this
to a future investigation.

Another interesting feature revealed by Fig.4 and Fig.5 is that, the
magnetization relaxation rate rises up quickly and monotonically to
100\% when the magnetic field is beyond the SP position $H_{sp}$.
This may suggest that the high field region is dominated by the
plastic motion of vortices. This conclusion will be corroborated by
the detailed analysis based on the collective pinning model in next
section.

\begin{figure}
\includegraphics[width=8cm]{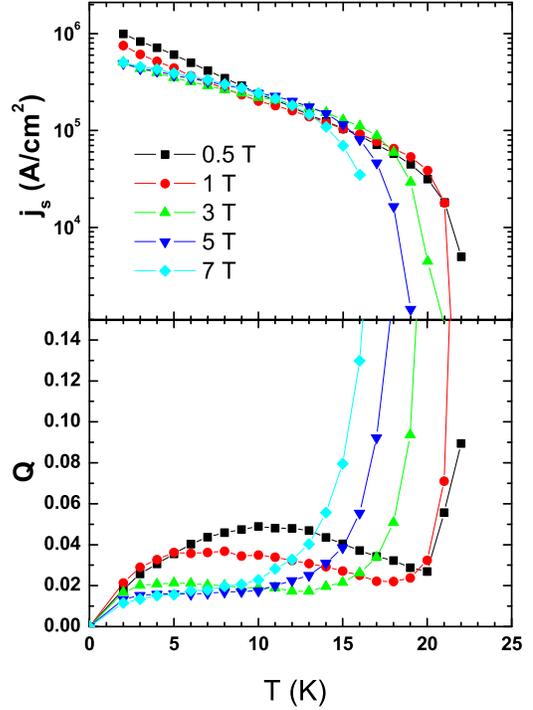}
\caption{(Color online) Temperature dependence of $\log j_s$ and
dynamic magnetization relaxation rate $Q$ with $x$ = 0.08 at 0.5, 1,
3, 5 and 7 Tesla. A plateau (or a weak bump) appears in the
intermediate temperature region.}\label{fig6 }
\end{figure}

\subsubsection{Analysis based on the vortex collective pinning model}

The temperature dependence of the transient superconducting current
density $j_{s}$ calculated through $\Delta M$ based on the Bean
critical state model and the dynamical magnetization relaxation rate
are presented in Fig.6. It is clear that the $\log j_s-T$ curve
shows a crossing feature at about 10 K for the data measured with
different magnetic fields, this is understandable since the SP
effect exhibits a crossover between different pinning regimes.
Interestingly the relaxation rate $Q$ shows a weak temperature
dependent or a bump-like behavior in the intermediate temperature
region. Although the relaxation rate $Q$ exhibits a rather large
value in the low temperature approach, we cannot conclude whether
the quantum tunneling of vortices is strong or not in the pnictide
superconductors since the lowest temperature measured here is about
2 K. The plateau or bump-like temperature dependence of $Q$ was also
observed in cuprate superconductors, which is especially pronounced
in YBa$_2$Cu$_3$O$_{7-\delta}$ superconductor.\cite{Fruchter} This
plateau cannot be understood within the picture of single vortex
creep with the rigid hopping length as predicted by the Anderson-Kim
model, but was attributed to the effect of collective pinning. We
will illustrate this point in the following discussion. In the high
temperature region, the relaxation rate rises sharply corresponding
to the plastic motion of vortices.

\begin{figure}
\includegraphics[width=8cm]{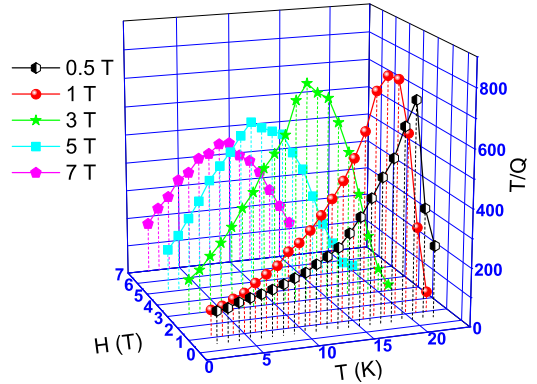}
\caption{(Color online) Temperature dependence of the ratio $T/Q(T)$
for the optimally doped sample for different fields. The slope of
$T/Q$ is positive in the low and intermediate temperature region,
indicative of an elastic vortex motion(see next). While it becomes
negative in the high temperature region revealing the plastic motion
of vortices.}\label{fig7}
\end{figure}

In order to understand the vortex motion in a detailed way, we
adopt the model of thermally activated flux motion (TAFM), which
reads

\begin{equation}
E=v_0B \exp (-\frac{U(j_s,T,B_e)}{k_BT}),
\end{equation}
here $E$ is the electric field induced by the vortex motion, $U(j_s,
T, B_e)$ is the activation energy, $v_0$ is the attempting hopping
velocity, $B_e$ is the actual local magnetic induction. It was
suggested that the activation energy can be written in a very
general way as

\begin{equation}
U(j_s,T,B_e)=\frac{U_c(T,B_e)}{\mu(T,B_e)}[(\frac{j_c(T,B_e)}{j_s(T,B_e)})^{\mu(T,B_e)}-1],
\end{equation}
where $\mu$, $U_c$, and $j_c$ are the glassy exponent, intrinsic
characteristic pinning energy and the un-relaxed critical current
density, respectively. The latter two parameters $U_c$ and $j_c$
strongly depend on the pinning details, such as the characteristics
of the pinning centers, the disorder landscape, the condensation
energy, the coherence length and the anisotropy, etc.. The glassy
exponent $\mu$ gives influence on the current dependence of $U$,
which is a decreasing function of $j_s$ for all possible values of
$\mu$. From the elastic manifold theory, it was predicted that $\mu$
= 1/7, 3/2, 7/9 for the single vortex, small bundles and large
bundles of vortex motion with the weak collective pinning
centers.\cite{VinokurPRL} For $\mu$ = -1 the equation recovers the
Kim-Anderson model, and the Zeldov logarithmic model as special
cases can be described for $\mu$=0.\cite{Zeldov} Just by combining
above two equations, it was derived that\cite{WenHHPhysicaC2001}

\begin{equation}
\mu = -Q \frac{\textrm{d}^2\ln E}{\textrm{d}\ln j_s^2}.
\end{equation}

This equation indicates that a negative $\mu$ would correspond to a
positive curvature of $\textrm{d}\ln E$ vs. $\textrm{d}\ln j_s$,
indicating of a finite dissipation in the small current limit and
plastic vortex motion, while a positive $\mu$ corresponds to a
negative curvature of $\textrm{d}\ln E$ vs $\textrm{d}\ln j_s$,
showing a vanishing dissipation in the small current limit and
elastic vortex motion. Therefore Eq.5 is physically meaningful for
any value of $\mu$, both positive and negative, as well as 0. Eq.5
is most suitable for a consistent analysis when a transition occurs
in the type of pinning as reported in this paper. From the general
vortex motion equations (3) and (4) mentioned above and the
definition of $Q$ , Wen et al. derived the following
equation\cite{WenHHJAlloy}

\begin{equation}
\frac{T}{Q(T, B_e)}=\frac{U_c(T,B_e)}{k_B}+\mu(T,B_e)CT,
\end{equation}
where $C=\ln(2v_0B/l\textrm{d}B_e/\textrm{d}t)$ is a parameter that
is weakly temperature dependent, $l$ is the lateral dimension of the
sample. We thus present the $T/Q$ vs. $T$ at different magnetic
fields in Fig.7. It is clear that the curve $T/Q$ vs. $T$ gives a
positive curvature in the low and intermediate temperature region,
while a negative slope appears in the high temperature region. Upon
above discussion, the positive slope of $T/Q$ vs. $T$ would suggest
an positive glassy exponent $\mu$ and elastic vortex motion, and a
negative slope of $T/Q$ vs. $T$ may suggest a negative $\mu$
provided the temperature dependence of $U_c(T)$ is not strong. By
extrapolating the curve $T/Q$ down to zero temperature, one can
obtain the value of $U_c(0)$. The value of $U_c(0)/k_B$ at 0.5 T
calculated from Fig.7 is about 98 K, which is actually a small
value, implying a quite small characteristic pinning energy. The
$U_c(0)$ is about 300 K (at 0.5 T) in YBCO thin
films\cite{WenPhysicaC1995,GriessenPRL1994} but beyond 3000 K in
MgB$_2$.\cite{MgB2Uc} With the increase of magnetic field, the
$U_c(0)/k_B$ gradually decreases. While there is an upturn at 7
Tesla, where the SP effect sets in. By calculating the slope of
$T/Q$ vs. $T$ curve, we can get the value of $\mu C$ assuming that
$U_c(T)$ is not a strongly temperature dependent function. In the
intermediate temperature region, $C$ can be derived from the
formula\cite{WenPhysicaC1995}

\begin{equation}
-\frac{\textrm{d}\ln j_s}{\textrm{d}T}=-\frac{\textrm{d}\ln
j_c}{\textrm{d}T}+C\frac{Q}{T}.
\end{equation}

Here we found $C$ = 28.56 $\pm 0.66$. In this way, $\mu$ can also be
roughly estimated. At 0.5 Tesla, we obtained a positive value $\mu$
= 0.45, which indicates an elastic vortex motion. The value obtained
here by the rough estimate is also consistent with what we get
through a quantitative analysis based on the
Generalized-Inversion-Scheme (GIS).

\subsubsection{Generalized-Inversion-Scheme}
\begin{figure}
\includegraphics[width=8cm]{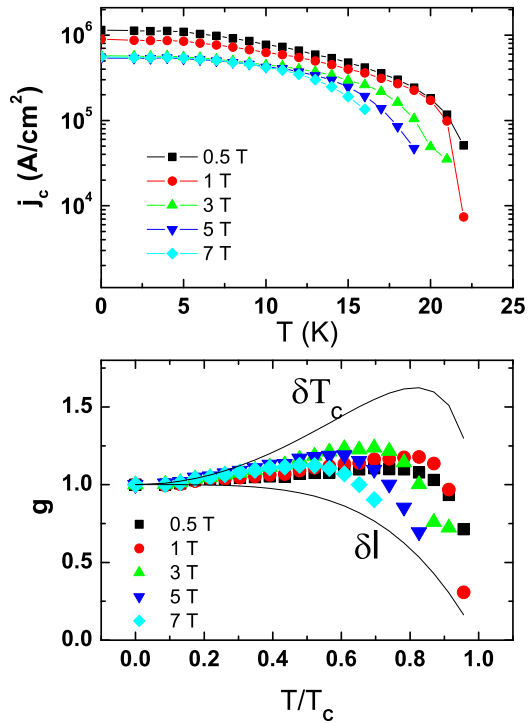}
\caption{(Color online) Temperature dependence of the unrelaxed
critical current density $j_c(T,B_e)$ and the normalized intrinsic
pinning energy $g(T, B_e)$ curves obtained by means of the GIS, for
the optimal doped sample. }\label{fig8}
\end{figure}

To extract information on the unrelaxed critical current density
$j_c(T,B_e)$ and the corresponding characteristic pinning energy
$U_c(T,Be)$ directly from the relaxation data, the GIS was proposed
by Schnack et al.\cite{Schnack} and Wen et
al.\cite{WenPhysicaC1995}. The basic assumptions of GIS are: (1)
TAFM [eq.(1)]; (2)$U(j_s,T)=U_c(0)g(t)f(j_s/j_c)$; (3)$g(t)\propto
[j_c(T)]^pG(T)$, here $U_c(0) g(j_s/j_c) G(t)$ and $p$ depend on the
specific pinning models (see below). These assumptions represent the
most general scheme among many methods proposed so far. According to
GIS, one can determine $j_c(T)$ by using the following integral:
\begin{equation}
\begin{split}
&j_c(T)= j_c(0)\times \exp[\\
&\int_{T}^{0}\frac{CQ(T')[1-\textrm{d}\ln G(T')/\textrm{d}\ln T']
+\textrm{d}\ln j_s(T')/\textrm{d}\ln T'}{1+pQ(T')C}\\
&\times \frac{\textrm{d}T'}{T'}].
\end{split}
\end{equation}

Here $j_c(0)$ is the critical current density at 0 K. In this
equation, the $Q(T)$ and $j_s(T)$ are the measured values. Once
$j_c(T)$ is known, the temperature dependence of the intrinsic
pinning energy $U_c(T)$ is obtained from assumption (3) of GIS. To
apply this procedure, one must know the value of $p$ as well as the
function $G(T)$. For a 2D pancake system, it is known that
$U_c(T)=j_c(T)\phi_0 d r_p$, where $\phi_0$ is the flux quanta, $d$
is the thickness of the pancake vortex, $r_p$ is the pinning range,
which typically is identified with the coherence length $\xi$. Thus,
in this special case, $p$ = 1 and $G(T)=\sqrt{(1+t^2)/(1-t^2)}$ with
$t=T/T_c$. Similarly, for a 3D single vortex, it is
found\cite{WenPhysicaC1995} that $p$ = 0.5 and
$G(T)=(1+t^2)^{5/4}/(1-t^2)^{1/4}$. By inserting these expressions
into above equation, one can calculate $j_c(T)$ and $U_c(T)$ from
the experimental data. The results of such a GIS analysis based on
the single vortex approach is presented in Fig.8(a) and (b). To be
consistent with the analysis given in the context of Fig.7, the same
value of $C$= 28 has been assumed. In Fig.8(b), we show together the
theoretical predictions for the two basic pinning mechanism: pinning
due to the spatial fluctuation of superconducting transition
temperatures $T_c$ which is called $\delta
T_c$-pinning,\cite{WenPhysicaC19952} and that due to the spatial
fluctuation of the mean-free-path, the so-called $\delta
l$-pinning.\cite{GriessenPRL1994,Blatter} Taking $p$ = 0.5 and $C$ =
28, the unrelaxed critical current density $j_c(T)$ and function
$g(T)$ are determined and presented in Fig.8. Shown together with
$g(T)$ are the theoretical predictions for the single vortex of
$\delta l$-pinning with $g(T) = 1-t^4$ and for the $\delta
T_c$-pinning $g(T) = (1-t^2)^\frac{2}{5}(1+t^2)^\frac{1}{2}$. One
can see that the experimentally derived value resides in between the
$\delta T_c$-pinning and $\delta l$-pinning. However an enhancement
of $g(t)$ in the high temperature region is hardly achieved by the
$\delta l$-pinning, which is however anticipated by the model of
$\delta T_c$-pinning. We would therefore conclude that either
$\delta T_c$-pinning or some other pinning mechanism are in
functioning in the present optimally doped samples. This of course
warrants further clarification.

\begin{figure}
\includegraphics[width=8cm]{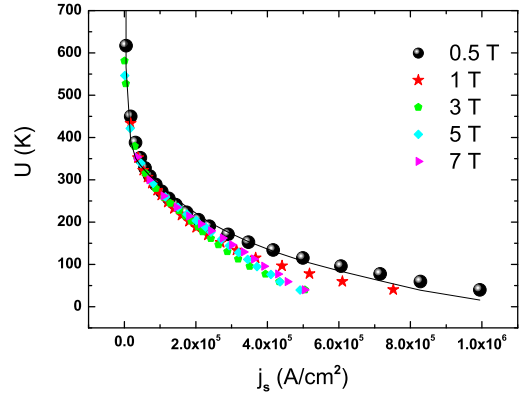}
\caption{(Color online) The $U(j_s, T, B_e)$ relation obtained by
means of the GIS with $p$ = 0.5, for the optimally doped sample. The
solid line gives a theoretical fit to the $U(j_s, T, B_e = 0.5T)$ as
formulated by eq.3 with $U_c$ = 116 K and $\mu=0.55 $. }\label{fig9}
\end{figure}

Fig.9 shows the $U(j_s, T, B_e)$ obtained by GIS at five fields for
the optimally doped sample. The data at 0.5 T was fitted to eq.4.
The fitting parameter $U_c$ is 116 K, which is close to 98 K
mentioned above. From the fit we get $\mu$ = 0.55, which is quite
close to that obtained before $\mu$ = 0.46. We should mention that
the $\mu$ value determined here is just an averaged one, which in
principle is also current dependent. Our work suggests that the
collective pinning model is applicable in this kind of
superconductors at low and intermediate temperature region with a
positive glassy exponent. At high fields (still below $H_{sp}$), the
collective pinning model may still work, but it is difficult to be
quantitatively described by the GIS since crossover from the single
vortex creep to small bundles or large bundles have occured.

\subsubsection{The vortex phase diagram}

\begin{figure}
\includegraphics[width=8cm]{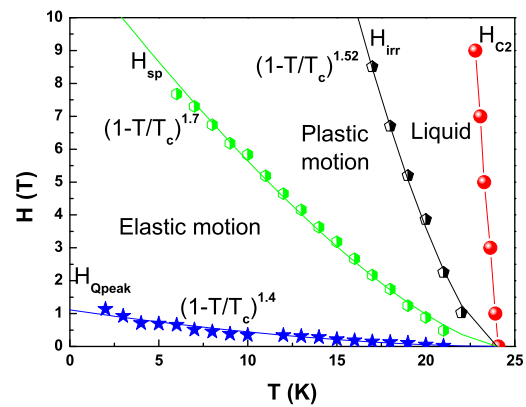}
\caption{(Color online) The vortex phase diagram of the optimally
doped sample with $x$ = 0.08. }\label{fig10}
\end{figure}

In Fig. 10, we present the vortex phase diagram of the optimally
doped sample with $x$= 0.08. From the magnetic measurement, three
characteristic fields are determined as shown by the solid symbols
in Fig.10. The second-peak $H_{sp}$ and the relaxation rate peak
$H_{Qpeak}$ together with the irreversibility field $H_{irr}$ (taken
with the criterion of 0.1 emu/cm$^3$) are shown. The upper critical
field $H_{c2}$ with 95 percent $\rho_n$ is shown by the filled
circles.
 There is a large area between the $H_{sp}-T$, and
$H_{irr}-T$ curves, suggesting that the vortex dissipation is
through a plastic motion in this region, but the dissipation level
is still quite low. The $H_{sp}-T$, $H_{Qpeak}-T$ and $H_{irr}-T$
are clearly temperature dependent. As an accumulation of knowledge,
the three curves are well fitted by the expressions
$H_{sp}(T)=H_{sp}(0)(1-T/T_c)^{1.7}$,
$H_{Qpeak}(T)=H_{Qpeak}(0)(1-T/T_c)^{1.4}$ and
$H_{irr}(T)=H_{irr}(0)(1-T/T_c)^{1.52}$, respectively.

\subsection{Overdoped sample with $x$=0.12}

\begin{figure}
\includegraphics[width=8cm]{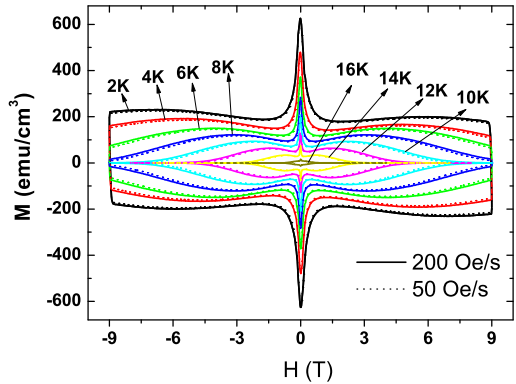}
\caption{(Color online) The MHLs of the overdoped sample ($x$ =
0.12) from 2 K to 16 K. The solid line represents the data measured
at the magnetic field sweeping rate of 200 Oe/s while dotted line at
50 Oe/s.}\label{fig11}
\end{figure}

Now we turn to the case of an overdoped sample ($x$ = 0.12) with
superconducting transition temperature of 19 K. The MHLs of the
sample $x$ = 0.12 measured at 2-16 K with the magnetic sweeping rate
50 Oe/s and 200 Oe/s are shown in Fig.11. At 2 K, the SP effect is
obviously observed, and by increasing temperature, the second-peak
shifts to the low magnetic field. The very sharp magnetization peak
was also found near the zero field. The symmetric curves suggest
that the bulk pinning instead of the surface barrier dominates in
the sample.

\begin{figure}
\includegraphics[width=8cm]{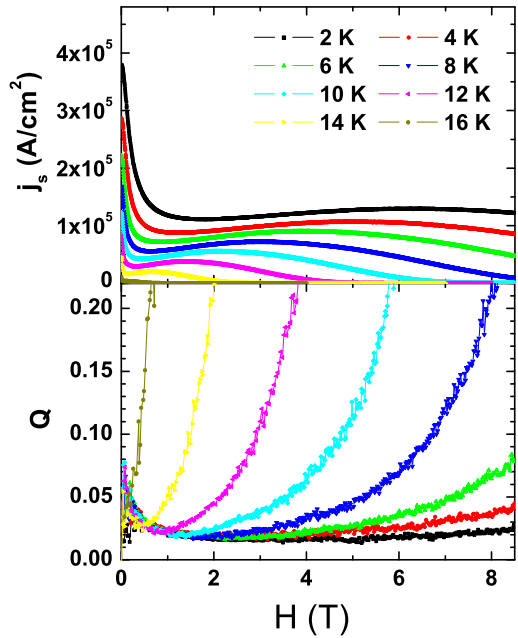}
\caption{(Color online) Magnetic field dependence of the transient
superconductivity current density is defer from the width
$\bigtriangleup M$ and dynamic magnetization relaxation rate from 2
K to 16 K with the sample $x$= 0.12. }\label{fig12}
\end{figure}

Based on the Bean critical state model,\cite{C. P. Bean} it is known
that the superconducting transient current $j_s\propto
\bigtriangleup M$. The non-monotonic field and temperature
dependence of $j_s$ was clearly observed in the SP region (shown in
Fig.12). At a low field, magnetization relaxation rate has an upturn
when decreasing the magnetic field, which is in accord with the
valley of the MHL. In the intermediate field region, $Q$ decreases
with the increasing of the magnetic field, this indicates a
crossover to a vortex glass region with relatively strong pinning of
vortex. Above the second-peak maximum value $H_{sp}$, the value of
$Q$ increases drastically showing a crossover to the regime of
plastic motion.

\begin{figure}
\includegraphics[width=8cm]{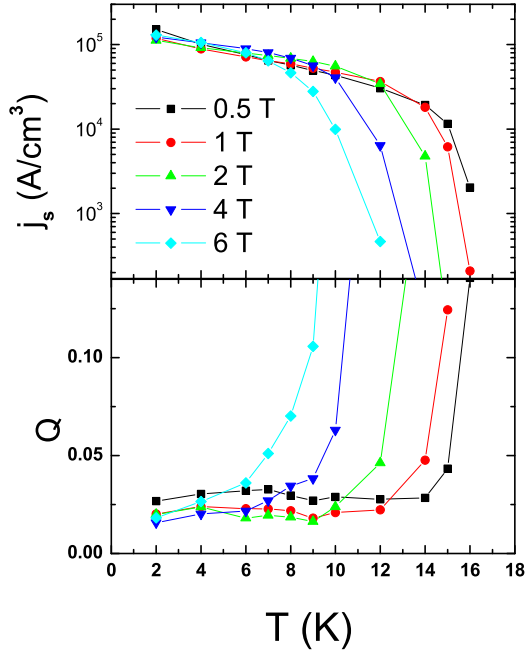}
\caption{(Color online) Temperature dependence of $\log j_s$ and the
dynamic magnetization relaxation rate $Q$ with $x$ = 0.12 at 0.5, 1,
2, 4 and 6 T. }\label{fig13}
\end{figure}

The temperature dependence of $Q$ and $j_s$ for the overdoped sample
$x$ = 0.12 was presented in Fig.13. Obviously, there is still a
plateau in the intermediate temperature region for each $Q-T$ curve,
which is corresponding to the linear region of $\log j_s-T$ curve.
And the region of the plateau shrinks quickly with increasing the
magnetic field. This behavior can be understood within the picture
of collective vortex pinning model. In the high temperature region,
the $Q$ increases abruptly, and the $\log j_s$ drops down
drastically, which again shows a crossover from the collective
elastic motion to plastic motion.

\begin{figure}
\includegraphics[width=8cm]{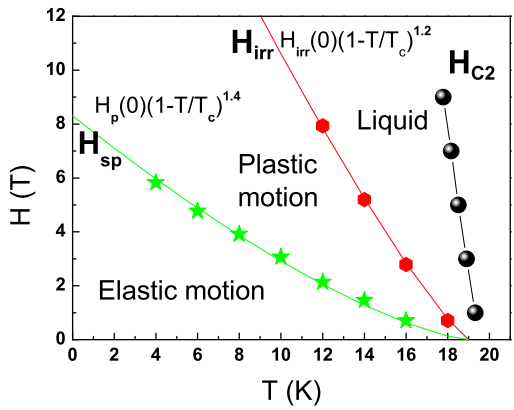}
\caption{(Color online) The vortex phase diagram of the overdoped
sample ($x$ = 0.12). Three characteristic fields $H_{sp}$, $H_{irr}$
and $H_{c2}$ are shown.}\label{fig14}
\end{figure}

The similar vortex behavior with the optimally doped sample can be
derived from the vortex phase diagram for the sample with $x$ = 0.12
(shown in Fig. 14). From the magnetic measurements, three
characteristic fields are determined as shown by the filled symbols
in Fig. 14. The SP field $H_{sp}$ locating at the peak of the
$j_s-H$ curve shown as the solid pentagram now is much lower than
that in the optimally doped sample. The irreversibility field
$H_{irr}$ is determined by taking a criterion of 0.1 emu/cm$^3$
shown as the solid star. The upper critical field $H_{c2}$ with 95
percent $\rho_n$ is shown as the filled circles. The $H_{sp}-T$,
$H_{irr}-T$ curves are well fitted by the expressions
$H_{sp}(T)=H_{sp}(0)(1-T/T_c)^{1.4}$ and
$H_{irr}(T)=H_{irr}(0)(1-T/T_c)^{1.2}$. Below the $H_{sp}$ line the
vortex behavior is consistent with the collective creep model.
Between $H_{sp}$ line and $H_{irr}$ line, it can be understood by
the model of plastic motion. Above the $H_{irr}$ line, the vortex
liquid phase exists.

\subsection{Underdoped sample with $x$ = 0.06}

\begin{figure}
\includegraphics[width=8cm]{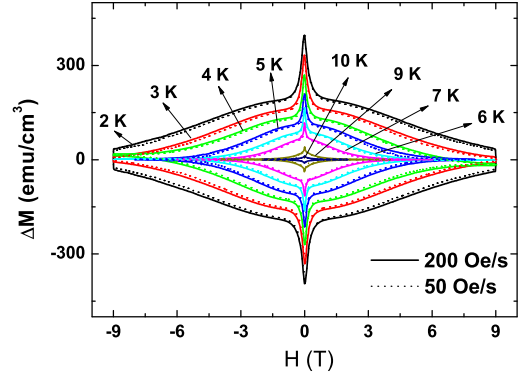}
\caption{(Color online) The MHLs of underdoped sample ($x$ = 0.06)
from 2 K to 10 K. The solid line represents the data measured at the
magnetic sweeping rate of 200 Oe/s while the dotted line at 50
Oe/s.}\label{fig15}
\end{figure}

Fig.15 shows the magnetization hysteresis loops of the sample $x$ =
0.06 measured at 2-10 K with the magnetic field sweeping rate 50
Oe/s and 200 Oe/s, respectively. At all temperatures we measured
here, the SP effect cannot be obviously observed. As the optimal
doped and overdoped sample the very sharp magnetization peak was
also found near the zero field. The symmetric MHL curves suggest
that the bulk pinning instead of the surface barrier dominates even
in this very underdoped sample.

\begin{figure}
\includegraphics[width=8cm]{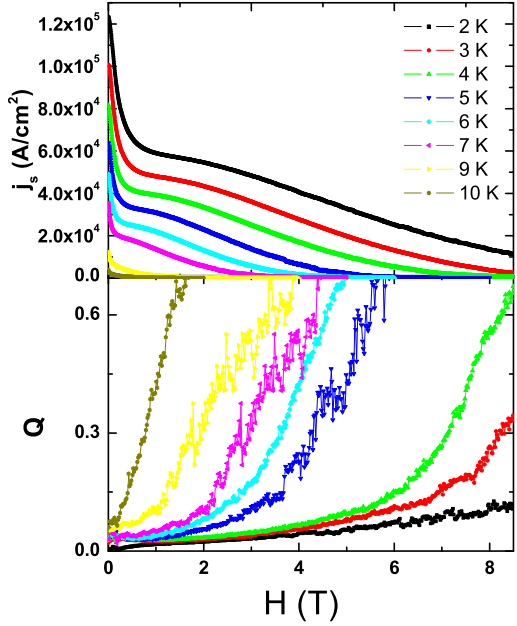}
\caption{(Color online) Magnetic field dependence of the MHL width
$\bigtriangleup M$ and dynamic magnetization relaxation rate
measured for the underdoped sample $x$=0.06 from 2 K to 10 K.
}\label{fig16}
\end{figure}

\begin{figure}
\includegraphics[width=8cm]{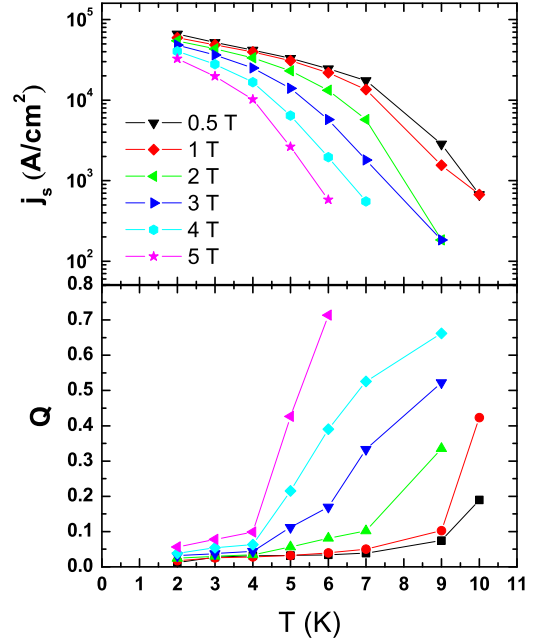}
\caption{(Color online) Temperature dependence of $\log j_s$ and
dynamic magnetization relaxation rate $Q$ with $x$ = 0.06 at 0.5, 1,
2, 3, 4 and 5 T. It is clear that the plateau of the relaxation rate
becomes very short followed by a quick rising up of relaxation
rate.}\label{fig17}
\end{figure}

The field dependence of $j_s$ and $Q$ from 2 K to 10 K was presented
in Fig.16. The $j_s$ decreases monotonously with the increase of
magnetic field, without showing a SP effect. Correspondingly, $Q$
increases with the increasing of the magnetic field rapidly and
quickly reaches the upper limit 100\%. This indicates that the
vortex motion in the most temperature regime investigated for this
sample is dominated by plastic motion. The temperature dependence of
$j_s$ and $Q$ measured at different magnetic fields were presented
in Fig.17. By increasing temperature, initially $\log j_s$ decrease
linearly, while it quickly evolves into a quick drooping down in the
high temperature region. In the low field region (0.5 T and 1 T),
the relaxation rate Q rises relatively slowly with temperature in
low and intermediate temperature region, while at 2 T, 3 T, 4 T and
5 T, the $Q$ curve increases dramatically on warming, which suggests
the quick evolvement of the plastic motion of vortices.

\begin{figure}
\includegraphics[width=8cm]{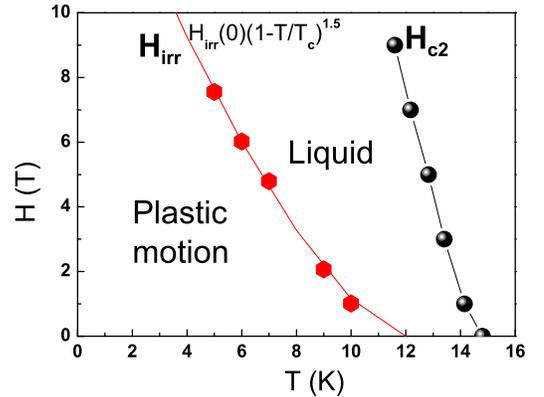} \caption{(Color online) The phase diagram of the ounderdoped sample
($x$=0.06) }\label{fig18}
\end{figure}

The vortex phase diagram of the underdoped sample with $x$ = 0.06 is
presented in Fig.18. Between the upper critical field and the
irreversibility line, it is the region for vortex liquid, while
there is now a region below the irreversible line which may have a
low dissipation, but be dominated by plastic motion. It is still
unclear whether in this sample we can still see the SP effect at
temperatures below 2 K. It is obvious that the SP effect is absent
 at temperatures above 2 K. One of the interpretations is that the elastic energy
which is based on the characteristic pinning energy is too weak to
sustain an elastic object in the vortex system. Therefore in this
sample, the vortex motion below $H_{irr}(T)$ is dominated by a
plastic manner. While we should not exclude the possibility that in
the regime at much lower temperatures, the elastic motion still
exists.

\section{Discussion}

\begin{figure}
\includegraphics[width=8cm]{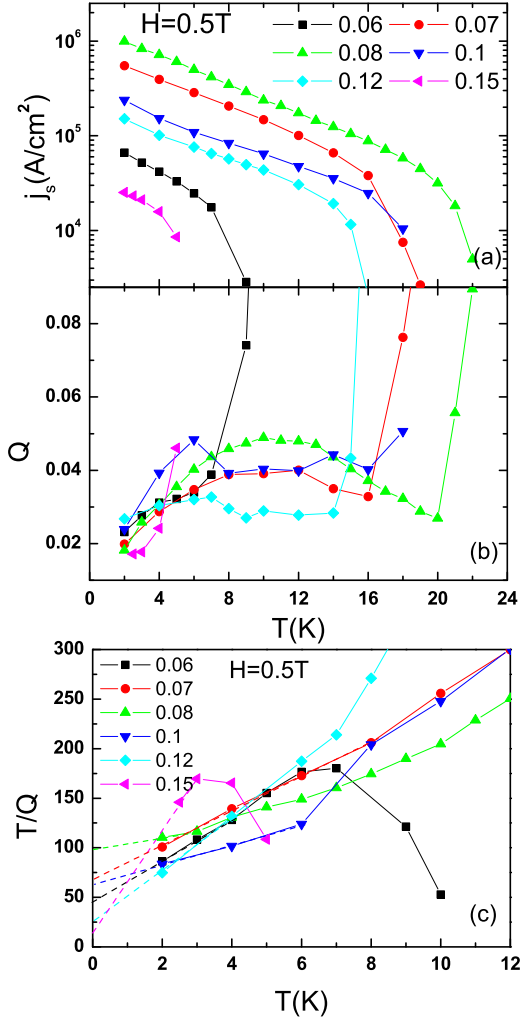} \caption{(Color online) Temperature dependence
of (a) the $\log j_s$, $Q$ and $T/Q$ for all six samples
investigated in this work. In low temperature region, a roughly
linear temperature dependence of $\log j_s(T)$ vs. T can be observed
for all samples besides the one with $x$ = 0.06.}\label{fig19}
\end{figure}

\begin{figure}
\includegraphics[width=8cm]{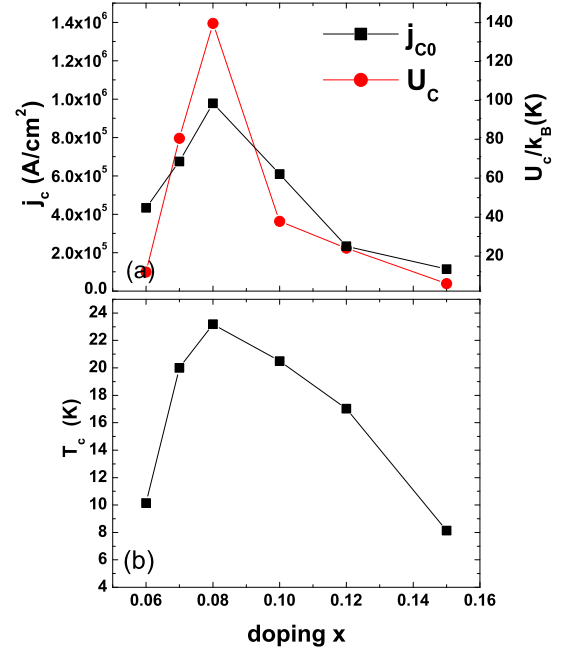} \caption{(Color online) Doping dependence of $j_{c0}$, $U_{c0}$
and $T_c$ for all six samples investigated here. It is clear that
both $j_{c0}$ and $U_{c0}$ have a very similar doping dependence of
$T_c$. }\label{fig20}
\end{figure}

\begin{figure}
\includegraphics[width=8cm]{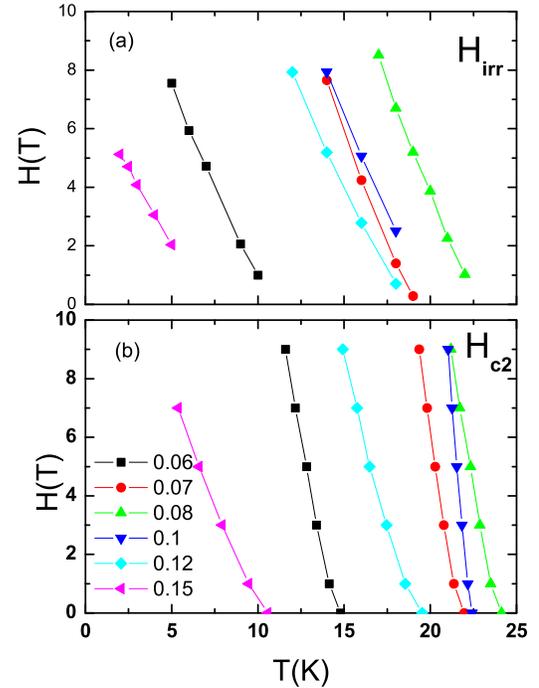} \caption{(Color online) Doping dependence of the $H_{irr}$
and $H_{c2}$ for all six samples.}\label{fig21}
\end{figure}

In all six samples investigated here, sizable magnetization
relaxation rate has been observed, which indicates a relatively
small characteristic pinning energy. In the samples around the
optimal doping level ($x \approx$ 0.08) a SP effect was easily
observed, while it becomes invisible in the very underdoped samples,
and hardly visible for the highly overdoped one. The missing of the
SP effect in the very underdoped and overdoped regions is
understandable since the pining energy becomes much weaker than that
of the optimally doped sample. In this case, the dislocation in the
vortex system can easily move leading to a plastic motion as
appeared in the high field region of the optimally doped sample.
Therefore we believe that the missing of the SP effect in the very
underdoped or overdoped samples is induced by the missing of the
elastic collective pinning and creep. The monotonic rising of the
relaxation rate in whole temperature region for the very underdoped
and overdoped samples support this argument. Another interesting
observation is that a very sharp magnetization peak was observed
near the zero field in all samples which is found to correspond very
well to a much reduced relaxation rate. Fig.19 shows the temperature
dependence of $Q$, $j_s$ and $T/Q$ with all doping levels at 0.5 T.
We found an bell-shaped or a plateau-like temperature dependence of
$Q$ with moderate relaxation rate in most samples, which can be
understood within the collective creep theory. The plateau region in
the very underdoped $x$ = 0.06 and very overdoped sample $x$ = 0.15
is very short because of the low pinning energy. According to eq.6,
the value of $T/Q$ in the $T$ = 0 K approach would give the
characteristic pinning energy $U_{c}(0)$. Fig.20 shows the doping
dependence of $U_c(0)$ at 0.5 T with all doping levels. The linear
$\log j_s$-T curves of all doped samples at 0.5 T suggest the
thermally activated collective vortex creep feature in the
intermediate temperature region. The $j_c(0)$, $U_c(0)$ and $T_c$
curves have the similar dome shape with doping, this may suggest
that the vortex pinning is through the disorders with small
condensation energy. By taking a GIS treatment on the magnetization
data, we found that the vortex pinning is probably achieved through
the spatial fluctuation of the superconducting transition
temperatures, the so-called $\delta$$T_c-$pinning. Further
investigations are strongly recommended to clarify this issue. With
the decreasing of $T_c$, the samples have lower irreversible field
and the upper critical field, as shown in Fig 21. While compared to
the underdoped samples, the overdoped samples with similar $T_c$
seem to have higher irreversible field and the upper critical field.
The SP in underdoped samples disappear more quickly with doping than
that in overdoped sample. For example, when the doping level drops
down from $x$ = 0.07 to 0.06, the SP effect disappears, while it is
still visible when the doping level goes to 0.12, although very
weak. This tendency is actually similar to the doping dependence of
$T_c$ and the characteristic pinning energy $U_c(0)$, as shown in
Fig. 20. This again suggests that the SP effect is dependent on the
pinning energy which governs actually the threshold of the plastic
motion of vortex system.

\section{Concluding Remarks}
We measured magnetization and its relaxation in
Ba(Fe$_{1-x}$Co$_x)_2$As$_2$ single crystals at various doping
levels ranging from very underdoped to very overdoped regime.
Detailed analysis lead to the following major conclusions:

(1) In all samples, sizable magnetization relaxation rate has been
observed, which suggests relatively weak vortex pinning energy. The
characteristic pinning energy obtained here for the optimally doped
sample is about 100 K at about 0.5 T.

(2) A very sharp magnetization peak was observed near zero field
which corresponds to a much reduced relaxation rate. This may be
induced by the extremely non-linear $B(x)$ near the edge when the
external field is swept to zero, or due to the surface barrier for
the vortex exit and entering at the edge.

(3) The second-peak effect was easily observed in the samples around
the optimal doping level ($x\approx$0.08), but it becomes hardly
visible in the very underdoped and highly overdoped samples. We
attribute the missing of the SP effect to the much weaker pinning
energy, which leads to a plastic motion of vortices in wide
temperature regions. Through the SP region the transient
superconducting current density shows the non-monotonic field and
temperature dependence.

(4) The weak temperature dependence of relaxation rate together with
the treatment of the Generalized-Inversion-Scheme, point to the fact
that the model of collective vortex pinning and creep works well in
describing the vortex dynamics in iron pnictides. The vortex pinning
is probably achieved through the spatial fluctuation of the
transition temperatures, which would mean an intrinsic inhomogeneity
of the iron-pnictide superconductors.

\section{Acknowledgments}

This work is supported by the Natural Science Foundation of China,
the Ministry of Science and Technology of China (973 project No:
2006CB60100, 2006CB921107, 2006CB921802), and Chinese Academy of
Sciences (Project ITSNEM).



Correspondence should be addressed to hhwen@aphy.iphy.ac.cn

\end{document}